\documentclass[aps,reprint,twocolumns,prl,superscriptaddress,floatfix,notitlepage,nofootinbib]{revtex4-1}
\usepackage{amssymb,amsmath,amsfonts} 
\usepackage{mathtools}
\usepackage{physics}
\usepackage{array}
\usepackage{graphicx,epsfig,xcolor}
\usepackage[colorlinks=true,citecolor=blue,linkcolor=red,pdftitle={equilibriumsymmetrybreaking},pdfauthor={CorpsRelano}]{hyperref}
\usepackage{newtxtext,newtxmath}
\usepackage{multibib}

\newcommand{\hatmath}[1]{\hat{\mathcal{#1}}} 
\newcommand{\beq}{\begin{equation}}
\newcommand{\eeq}{\end{equation}}

\begin{document}

\title{General theory for discrete symmetry-breaking equilibrium states}

\author{\'{A}ngel L. Corps}
    \email[]{corps.angel.l@gmail.com}
    \affiliation{Instituto de Estructura de la Materia, IEM-CSIC, Serrano 123, E-28006 Madrid, Spain}
    \affiliation{Grupo Interdisciplinar de Sistemas Complejos (GISC),
Universidad Complutense de Madrid, Av. Complutense s/n, E-28040 Madrid, Spain}
    
\author{Armando Rela\~{n}o}
    \email[]{armando.relano@fis.ucm.es}
    \affiliation{Grupo Interdisciplinar de Sistemas Complejos (GISC),
Universidad Complutense de Madrid, Av. Complutense s/n, E-28040 Madrid, Spain}
    \affiliation{Departamento de Estructura de la Materia, F\'{i}sica T\'{e}rmica y Electr\'{o}nica, Universidad Complutense de Madrid, Av. Complutense s/n, E-28040 Madrid, Spain}

\date{\today} 

\begin{abstract}
Spontaneous symmetry-breaking in phase transitions occurs when the system Hamiltonian is symmetric under a certain transformation, but the equilibrium states observed in nature are not. Here, we prove that when a discrete symmetry is spontaneously broken in a quantum system, then the time evolution necessarily conserves two additional and non-commuting quantities, besides the one linked to the symmetry. This implies the existence of equilibrium states consisting in superpositions of macroscopic configurations. Then, we propose an experimental realization of such equilibrium states with the current state-of-the art in quantum technologies. Through numerical calculations, we show that they survive as very long-lived pre-thermal states, even very far away from the thermodynamic limit. Finally, we also show that a small symmetry-breaking perturbation in the Hamiltonian stabilizes the conservation of one of the two former quantities, implying that symmetry-breaking equilibrium states become stable even in small quantum systems.
\end{abstract}

\maketitle

Spontaneous symmetry-breaking (SSB) is a cornerstone of many branches of physics, like phase transitions in condensed matter \cite{Landaubook}, non-equilibrium thermodynamics \cite{Kondepudi1983}, elementary particles \cite{Baker1962,Goldstone1962} and cosmology \cite{Kibble1976}. In the first case, it appears when the Hamiltonian is symmetric under a certain transformation, but equilibrium states observed in nature are not. Typically, this is manifested in the form of an order parameter which is either positive or negative in equilibrium, even though it is necessarily zero in symmetric states. The accepted explanation of this phenomenon relies on the singular nature of the thermodynamic limit (TL) \cite{Anderson1972,Berry2002}. When a symmetry-breaking perturbation is introduced into the Hamiltonian, the final result depends on whether the TL is taken before reducing this perturbation to zero, or afterwards. This implies that the effect of a tiny perturbation becomes frozen in the equilibrium state. Notwithstanding, SSB is always observed in {\em finite} systems, and thus an explanation based on the singularity of such infinite-size limit is not completely satisfactory. The aim of this article is to introduce a statistical ensemble accounting for all the dynamical consequences of breaking a discrete symmetry, and to show that it can be used to explain SSB even in small quantum systems.

\section{Dynamical consequences of discrete symmetry-breaking}

We start by assuming that symmetry-breaking equilibrium states do exist. From this fact, formulated in terms of four hypothesis about the mathematical properties of the system, we present four theorems allowing us to derive a statistical ensemble accounting for all possible kinds of equilibrium states within a discrete symmetry-breaking phase (proofs are given in Supplemental material).

The first hypothesis is:

\textbf{H1.-} There exists a $\mathbb{Z}_{2}$ symmetry, $\hat{\Pi}$, that we call \textit{parity}, labeling the Hamiltonian eigenstates as $\hat{\Pi}\ket{E_{n,\alpha}}=\alpha\ket{E_{n,\alpha}}$, $n=0,1,2,...$, $\alpha\in\{+,-\}$, where $\hat{H}\ket{E_{n,\alpha}}=E_{n,\alpha}\ket{E_{n,\alpha}}$. 

Equilibrium states for which this discrete symmetry is broken are usually identified by an order parameter, $\hat{M}$, whose expectation value is zero in any symmetric state. A sufficient condition for this is
\beq\label{eq:propertyM}
\hat{M}\ket{E_{n,\alpha}}=\sum_{m}c_{m}\ket{E_{m,-\alpha}}.
\eeq
Hence, we rely on this kind of observables to formulate the next two hypothesis:

\textbf{H2.1.-} There exists a subspace $\mathcal{H}_D \subset \mathcal{H}$, where $\mathcal{H}$ is the Hilbert space of the system, spanned by a number of eigenstates of the Hamiltonian, $\{\ket{E_{m,\alpha}}\}$, such that for all $\ket{\psi}\in\mathcal{H}_{D}$, the time evolution $\ket{\psi(t)}=e^{-i\hat{H}t}\ket{\psi}$ verifies
\beq
\label{eq:mD}
\overline{\langle \hat{M}\rangle}\equiv \lim_{\tau\to\infty}\frac{1}{\tau}\int_{0}^{\tau}\bra{\psi(t)}\hat{M}\ket{\psi(t)}\textrm{d}t=0.
\eeq

\textbf{H2.2.-} $\mathcal{H}_{D}$ is the largest subspace spanned by the eigenstates of the Hamiltonian satisfying (H2.1).

\textbf{H2.3.-} $\mathcal{H}_{D}\neq\mathcal{H}$, i.e., there exists $\mathcal{H}_{SB}\subset\mathcal{H}$ such that $\mathcal{H}=\mathcal{H}_{D}\oplus\mathcal{H}_{SB}$. This entails the existence of states $\ket{\phi} \in \mathcal{H}_{SB}$ such that
\beq
\label{eq:mSB}
\overline{\langle \hat{M}\rangle}\equiv \lim_{\tau\to\infty}\frac{1}{\tau}\int_{0}^{\tau}\bra{\phi(t)}\hat{M}\ket{\phi(t)}\textrm{d}t \neq 0.
\eeq
where $\ket{\phi(t)}=\textrm{e}^{-i\hat{H}t} \ket{\phi}$. 

As infinite-time averages, like those of Eqs. \eqref{eq:mD} and \eqref{eq:mSB}, represent equilibrium states in isolated quantum systems \cite{Reimann2008}, $\mathcal{H}_{SB}$ accounts for the symmetry-breaking phase, and it is typically spanned by all the Hamiltonian eigenstates with associated eigenenergies below the one corresponding to the critical temperature of the phase transition, $T_c$. On the other hand, $\mathcal{H}_D$ accounts for the disordered phase.

From these two hypothesis, we formulate the following theorem.

\textbf{Theorem 1.} \textit{Hypothesis (H1) and (H2) hold true if and only if there exist three operators, $\hat{O}_{1}$ $\hat{O}_{2}$ and $\hat{O}_{3}\equiv \hat{\Pi}$ verifying:}

\textit{(i) The set $\{\hat{O}_{1},\hat{O}_{2},\hat{O}_{3}\}$ is a set of $\mathbb{Z}_{2}$ operators: $\hat{O}_{1}^{2}=\hat{O}_{2}^{2}=\hat{O}_{3}^{2}=\mathbb{I}$.}

\textit{(ii) The set $\{\hat{O}_{1},\hat{O}_{2},\hat{O}_{3}\}$ satisfy the SU(2) commutation algebra: $[\hat{O}_{m},\hat{O}_{n}]=2i\varepsilon_{mn\ell}\hat{O}_{\ell}$, where $\varepsilon_{mn\ell}$ is the Levi-Civita symbol.}

\textit{(iii) The set $\{\hat{O}_{1},\hat{O}_{2},\hat{O}_{3}\}$ commute with the projection of $\hat{H}$ onto the subspace $\mathcal{H}_{SB}$, but $\hat{O}_1$ and $\hat{O}_2$ do not commute with the projection of $\hat{H}$ onto the subspace $\mathcal{H}_D$.} 

Theorem 1 implies that {\em unitary dynamics in the symmetry-breaking phase is qualitatively different from unitary dynamics in the disordered phase.} In the first case, $\langle \hat{O}_1 \rangle$, $\langle \hat{O}_2\rangle$, and $\langle \hat{O}_3\rangle$ remain constant, and therefore any statistical ensemble must depend on these three expectation values to properly describe equilibrium states. Yet, in the disordered phase all information about the initial values of $\langle \hat{O}_1\rangle$ and $\langle \hat{O}_2\rangle$ is effectively erased by the unitary evolution, and hence only $\langle \hat{O}_3 \rangle$ is required to build a proper equilibrium ensemble.

The main drawback of this theorem is that it does not specify the form of $\hat{O}_1$, $\hat{O}_2$, and $\hat{O}_3$. To cover this gap, we also require:

\textbf{H3.-} There exist a subspace $\mathcal{H}_{\mathcal{C}}$, spanned by a number of Hamiltonian eigenstates, and an operator $\hat{A}$ such that, if the probability of initially obtaining a value $A>0$ (or $A<0$) in a measurement is zero, then this property is conserved at all subsequent times by the Hamiltonian dynamics within $\mathcal{H}_{\mathcal{C}}$. This means that, within this subspace, there exists two distinct {\em wells}, defined by the sign of the eigenvalues of $\hat{A}$. 

This hypothesis accounts for a typical behavior in symmetry-breaking phase transitions: when the system is cooled below the critical temperature, the time-evolving state is trapped in one of the two parts of a double well, one characterized by positive values of the order parameter, and the other one characterized by negative values. Note, however, that (H3) is more general, since it does not require the existence of such a phase transition. We will see later that this makes it possible to define two different {\em dynamical phases} in systems including a symmetry-breaking term annihilating the critical behavior.

From this hypothesis, we formulate:

\textbf{Theorem 2.1.} \textit{(H3) holds true if and only if the operator $\hat{\mathcal{C}}=\textrm{sign}\,(\hat{A})$ is constant within $\mathcal{H}_{\mathcal{C}}$. }

If we also assume (H1) and (H2), and $\hat{A}=\hat{M}$ in \eqref{eq:propertyM},  then (H3) is more restrictive than (H1)-(H2): (H3) implies the existence of initial states for which $\overline{\langle \hat{M}\rangle}\neq 0$, but the converse does not hold true. This implies that $\mathcal{H}_{\mathcal{C}} = \mathcal{H}_{SB}$.

Then, we formulate:

\textbf{Theorem 2.2.} \textit{(H1)-(H3) hold true if and only if the operators $\hat{\mathcal{C}}=\textrm{sign}\,(\hat{M})$ and $\hat{\mathcal{K}}=\frac{i}{2}[\hat{\mathcal{C}},\hat{\Pi}]$ are constant in the Hilbert subspace $\mathcal{H}_{SB}$ of Theorem 1.}

To fully achieve our goal of identifying the operators $\hat{O}_1$, $\hat{O}_2$, and $\hat{O}_3$, we need a link between them and the operators $\hat{\mathcal{C}}$ and $\hat{\mathcal{K}}$ of Theorem 2.2. This is obtained from:

\textbf{Theorem 3.} \textit{Suppose (H1)-(H3) hold true. Then, the operators $\hat{\mathcal{C}}=\textrm{sign}(\hat{M})$ and $\hat{\mathcal{K}}=\frac{i}{2}[\hat{\mathcal{C}},\hat{\Pi}]$ of Theorem 2.2 coincide with the operators $\hat{O}_{1}$ and $\hat{O}_{2}$ of Theorem 1.} 

The main physical consequence of these theorems is that $\langle \hat{\mathcal{C}}\rangle$, $\langle \hat{\mathcal{K}}\rangle$, and $\langle \hat{\Pi} \rangle$ are necessary to describe equilibrium states within the symmetry-breaking phase of any system fulfilling hypothesis (H1)-(H3). It follows from Theorem 2.2 that, if $\langle \hat{\mathcal{C}}\rangle =1$, then the corresponding state is trapped in the double-well region where the probability of obtaining a negative value for $\hat{M}$ is zero, and the opposite happens if $\langle \hat{\mathcal{C}}\rangle =-1$; so, $\langle \hat{\mathcal{C}}\rangle$ determines the probability of observing any of these two possibilities in a measurement of $\hat{M}$. Then, $\langle \hat{\Pi}\rangle$ and $\langle \hat{\mathcal{K}} \rangle$ determine the quantum coherence between these two possibilities. And therefore {\em equilibrium states consisting in coherent superpositions of states trapped in these two different wells may exist.}

Finally, we propose a statistical ensemble to account for this phenomenon. The starting point is our last hypothesis:

\textbf{H4.-} Let $\hat{\rho}=\frac{1}{Z}e^{-\beta \hat{H}}$ denote a canonical density matrix, and $\beta=(k_{B}T)^{-1}$ ($k_{B}=1$ hereinafter) be its inverse temperature. Then, the average energy $\langle E\rangle_{\rho}=\Tr[\hat{\rho} \hat{H}]$ is an extensive quantity, and the quantum fluctuations around this value vanish in the thermodynamic limit, $\left(\frac{\sigma_{E}}{\langle E\rangle}\right)_{\rho}\to0$ as $N\to \infty$ (this is typically fulfilled by realistic systems). 

From this hypothesis, we formulate:

\textbf{Theorem 4.} \textit{If a physical system satisfies (H1)-(H4), then the density matrix}
\beq\label{eq:rhoGGE}
\hat{\rho}_{\textrm{GGE}}=\frac{1}{Z}e^{-\beta\hat{H}-\lambda_{c}\hat{\mathcal{C}}-\lambda_{k}\hat{\mathcal{K}}-\lambda_{\pi}\hat{\Pi}}
\eeq
\textit{where $Z$ ensures that} $\textrm{Tr} [\hat{\rho}_{\textrm{GGE}}]=1$, \textit{is a constant of motion below the critical temperature of the symmetry-breaking phase transition,}
\beq
[\hat{H},\hat{\rho}_{\textrm{GGE}}]=0,\,\,\,\beta>\beta_{c},\,\,\,\forall \lambda_{\pi},\lambda_{c},\lambda_{k},
\eeq

Theorem 4 is our main result. It implies that $\hat{\rho}_{\textrm{GGE}}$, given by Eq. \eqref{eq:rhoGGE}, is an equilibrium ensemble below the critical temperature of the phase transition. The corresponding density matrix maximizes the entropy, conditioned by the constraints imposed by the conservation of energy, $\hat{\Pi}$, $\hat{\mathcal{C}}$ and $\hat{\mathcal{K}}$ in the symmetry-breaking phase \cite{Jaynes1957I,Jaynes1957II}. It has the shape of a generalized Gibbs ensemble (GGE), irrespective of whether the system is integrable or not \cite{Rigol2007}, composed by a set of non-commuting charges \cite{Fagotti2014,Guryanova2016,Halpern2016}. 
$\beta$ determines the temperature, and $\lambda_c$, $\lambda_k$ and $\lambda_{\pi}$, the values of $\langle \hat{\mathcal{C}}\rangle$, $\langle \hat{\mathcal{K}} \rangle$ and $\langle \hat{\Pi}\rangle$, which remain constant. Above the critical temperature, as neither $\langle \hat{\mathcal{C}} \rangle$, nor $\langle \hat{\mathcal{K}}\rangle$ are conserved charges, equilibrium states are given by Eq. \eqref{eq:rhoGGE} with $\lambda_c = \lambda_k=0$.

$\hat{\rho}_{\textrm{GGE}}$ accounts for all of the different equilibrium states that can be observed in a symmetry-breaking phase, and allows us to classify them in three different families:

\textbf{ES1.-} Typical symmetry-breaking states. In them, the system remains trapped in one of the two wells of the order parameter, giving rise to either $M>0$ or $M<0$ in any measurement of $\hat{M}$. They are described by Eq. \eqref{eq:rhoGGE} with $\lambda_k = \lambda_{\pi}=0$, and $\lambda_c \rightarrow \pm \infty$. Note that {\em they cannot be obtained by means of the standard Gibbs ensemble}, which is recovered from Eq. \eqref{eq:rhoGGE} with $\lambda_c=\lambda_k=\lambda_{\pi}=0$. 

\textbf{ES2.-} Statistical mixtures of the typical symmetry-breaking states. This possibility is given by $\lambda_k = \lambda_{\pi}=0$, and $\left| \lambda_c \right| < \infty$; the statistical weight of each well is given by $\lambda_c$. 

\textbf{ES3.-} Macroscopic superpositions of the two wells of the order parameter.
They are given by $\lambda_k \neq 0$ and/or $\lambda_{\pi} \neq 0$. Thus, 
$\lambda_k$ and $\lambda_{\pi}$ determine the quantum coherence between these two wells.

\section{Implementation with cold atoms}  

The previous theorems are derived assuming the existence of symmetry-breaking equilibrium states. But it is well known that the conditions required for their existence are only fulfilled in the TL. Hence, our next step is to propose a protocol to study their practical consequences relying on the state-of-the art techniques in quantum technologies. We have chosen the transverse-field Ising model (TFIM),
\begin{equation}\label{eq:TFIM}
    \hat{H}_{\textrm{TFIM}}=-\sum_{i,j} V_{ij}\hat{\sigma}_{i}^{x}\hat{\sigma}_{j}^{x}+h\sum_{i}\hat{\sigma}_{i}^{z},
\end{equation}
with long-range and ferromagnetic interactions, $V_{ij} \propto |i-j|^{-\alpha} >0$. In Eq. \eqref{eq:TFIM}, $\hat{\sigma}_{i}^{x,y,z}$ are the Pauli matrices acting on site $i$, and $h$ is the magnetic field. This Hamiltonian is invariant under a $180$ degrees rotation around the z-axis, and therefore its parity is given by, $\hat{\Pi}=\prod_j \hat{\sigma}_j^z$. The magnetization $\hat{M} = \sum_i \sigma_i^x$ is good order parameter satisfying Eq. \eqref{eq:propertyM}. This model has been recently used to study the dynamical consequences of crossing its quantum critical point \cite{Islam2013,Islam2011,Simon2011,Britton2012}. For $\alpha < 2$, there exists a symmetry-breaking phase for $h<h_{c}$ and $T<T_{c}$ \cite{GonzalezLazo2021}. Here we propose a protocol similar to those of the previous references:

\textbf{S1.-} Start from a fully polarized state in the $X$ direction, $\ket{\uparrow \uparrow \cdots \uparrow \uparrow}_x$ or $\ket{\downarrow \downarrow \cdots \downarrow \downarrow}_x$. 

\textbf{S2.-} Activate an adiabatic ramp to slowly increase the transverse field from $h=0$ to $h=h_1 > h_c$.

\textbf{S3.-} Let the system relax at $h_1$ during a controlled time, $\tau_R$.

\textbf{S4.-} Activate a second adiabatic ramp to slowly decrease the transverse field from $h=h_1$ to $h=h_2 < h_c$.

\textbf{S5.-} Quench the system from $h=h_2$ to $h=h_3<h_2$.

As the initial state is a symmetry-breaking ground state of \eqref{eq:TFIM} with $h=0$ and $\langle\hat{\Pi}\rangle=0$, if step S2 is performed slowly enough, the unitary time evolution only introduces an irrelevant global phase until $h=h_c$. Then, as parity remains conserved, both the ground state ($\langle \hat{\Pi}\rangle=1$) and the first excited state ($\langle \hat{\Pi}\rangle=-1$) become equally populated. Hence, as S4 induces basically the same changes that S2 in the time-evolved wavefunction, the system would be prepared in a superposition of the two degenerate lowest-energy eigenstates of $\hat{H}(h_2)$, 
\beq \label{eq:initialstate} \ket{\Psi}=\sqrt{p}\ket{E_{0,+}}+e^{i\phi}\sqrt{1-p}\ket{E_{0,-}},
\eeq
with $p=1/2$ and an uncontrolled phase $\phi=\phi_q$ just before S5, if S3 was not done. The main consequence of the intermediate S3 is thus to introduce an extra {\em controlled} phase, $\phi_R$, in the state after S4. In terms of the gap $\Delta E_0=|E_{0,+}-E_{0,-}|$, at $ \hat{H}(h_1)$, this phase is simply $\phi_R=\Delta E_0 \, \tau_R$, so {\em a complete $2\pi$-period in the final phase $\phi=\phi_q + \phi_R$ can be explored by considering $0 \leq \tau_R < 2 \pi / \Delta E_0$}. Finally, S5 heats the system. If the final temperature is below $T_{c}$, then the initial expectation values of $\hat{\Pi}$, $\hat{\mathcal{C}}$ and $\hat{\mathcal{K}}$, given by
\begin{subequations}
\begin{eqnarray}
\label{eq:O1}
\langle \hat{\mathcal{C}}\rangle_{\Psi} &=& 2\sqrt{p(1-p)}\cos\phi, \\ 
\label{eq:O2}
\langle \hat{\mathcal{K}}\rangle_{\Psi} &=& 2\sqrt{p(1-p)}\sin\phi,  \\
\label{eq:O3}
\langle \hat{\Pi} \rangle_{\Psi} &=&2p-1,
\end{eqnarray}
\end{subequations}
remain constant. This means that we can prepare symmetry-breaking states with controlled values of $\langle\hat{\mathcal{C}}\rangle$, $\langle \hat{\mathcal{K}}\rangle$, and $\langle \hat{\Pi}\rangle=0$ by tuning $\tau_R$ appropriately.

The main inconvenience of this protocol is that the time-evolving state crosses a QPT twice. It is well known that this may induce uncontrolled excitations \cite{Kibble1976,Zurek1985,Zurek2005}. To estimate their importance, 
we take advantage of the fact that the QPT of the TFIM is in the same universality class as that of its fully-connected counterpart \cite{Dutta2001}, given by $V_{ij}=1/N$ in Eq. \eqref{eq:TFIM}. In this particular case, the Hamiltonian is commuting with the total angular momentum, so we work with the maximally symmetric sector, $j=N/2$, which includes the ground-state of the system. As a consequence, we can work with a much smaller Hilbert space.
 
\begin{figure}[h!]
\hspace{-0.7cm}\includegraphics[width=0.52\textwidth]{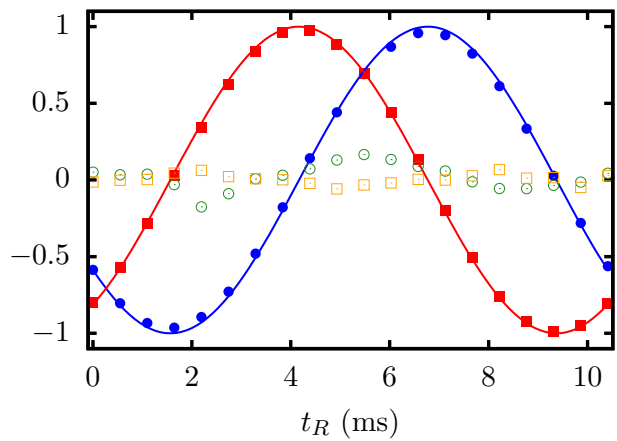}
    \caption{\textbf{Accuracy of the protocol.} Expectation values of the operators $\hat{\mathcal{C}}$ (circles) and $\hat{\mathcal{K}}$ (squares) after step S4, as a function of the relaxation time $t_{R}$. Filled markers are for $\tau_{q}=40.96$ ms and empty markers for $\tau_{q}=0.9$ ms. Solid lines represent the periodic result expected from a perfectly adiabatic protocol. System size is $N=20$.}
    \label{fig:sch}
\end{figure}

\begin{figure*}[t!]
\centering
    \begin{tabular}{c c}

    \hspace{-0.4cm}\includegraphics[width=0.52\textwidth]{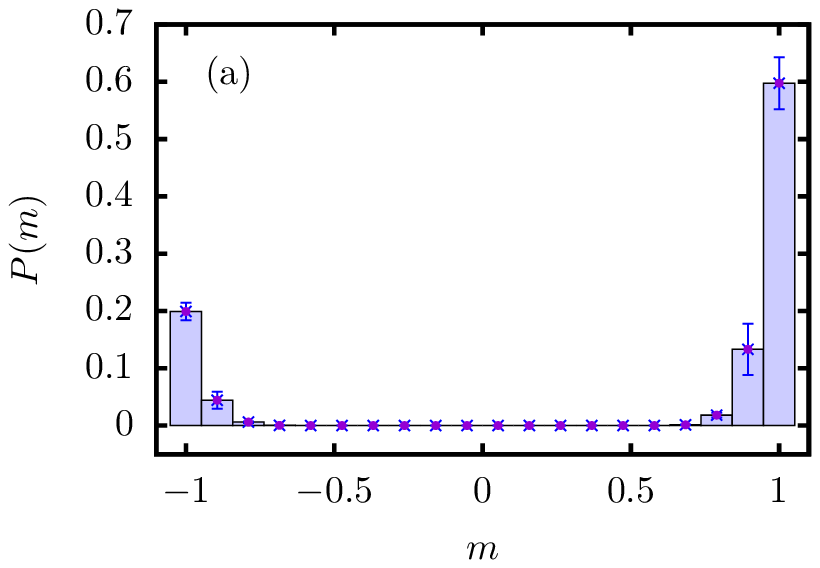} & 
\hspace{-0.7cm}\includegraphics[width=0.52\textwidth]{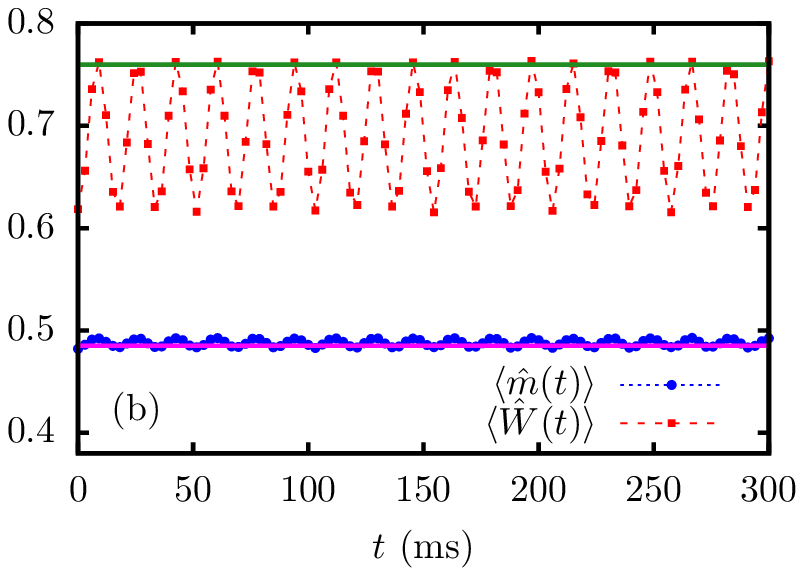}  \\
    \end{tabular}
    \caption{\textbf{Generation of equilibrium states.} (a) Distribution of the scaled collective magnetization, $m=M/N$, corresponding to a quench $h_{2}=0.5\to h_{3}=0.1$ in the TFIM with $\alpha=1.1$, $J=2$, and $N=19$. The initial state has $p=1/2$ and $\phi=\pi/3$. Errorbars represent the standard deviation of the instantaneous result from the mean over the entire time interval $t\in[0,300]$.
    (b) The instantaneous expectation value of the scaled collective magnetization $\hat{m}$ and $\hat{W}$ after the same quench. The predictions of the GGE for $\hat{m}$ (magenta) and $\hat{W}$ (green) are also shown with solid horizontal lines. }
    \label{fig:schandquench}
\end{figure*}

In Fig. \ref{fig:sch} we represent the value of $\hat{\mathcal{C}}$ and $\hat{\mathcal{K}}$ for two values of the driving time after the step S4, using an adiabatic ramp given by $h(t)=t/\tau_q$. These values are represented as a function of the relaxation time, $\tau_{R}$, spent by the state in $\hat{H}(h_{1})$. For a slow driving, $\tau_{q}=40.96$ ms (we use the time scales of the experiment in \cite{Islam2013}), we observe periodic oscillations in both observables; solid curves represent the theoretical behavior expected for a $2 \pi$-period in the relative phase $\phi$. 
On the other hand, for the fast driving protocol with $\tau_{q}=0.9$ ms, we observe that the previous oscillatory structure is lost. Of course, larger times are required to recover the $2 \pi$-period in the observables in larger systems \cite{Puebla2019}. Notwithstanding, it is worth noting that a similar protocol, crossing the QPT just once and relying on an exponential ramp, has been already performed \cite{Islam2013,Islam2011,Simon2011,Britton2012}.

Now, let us assume that we have successfully performed this preparation procedure with a long-range TFIM, obtaining a state given by Eq. \eqref{eq:initialstate} after step S4 with $p=1/2$ and $\phi=\pi/3$. Next, we numerically study the consequences of step S5. We work with a power-law interaction, $V_{ij} \propto J \left| i-j \right|^{-\alpha}$, with $\alpha=1.1$ and strength $J=2$, and we perform the final quench from $h=0.5$ to $h=0.1$ (see Supplemental material for more details).

In Fig. \ref{fig:schandquench}(a), we show the time-averaged probability of obtaining a value $m \in [-1,1]$ for the scaled magnetization, $\hat{m} = (1/N) \sum_i \hat{\sigma}^x_i$, in a measurement. We observe a highly asymmetric distribution, which it is not invariant under a $180$ degree rotation around the z-axis; therefore, we observe a symmetry-breaking time-averaged state. 

In Fig. \ref{fig:schandquench}(b), we show the time evolution $\left\langle \hat{m} (t) \right\rangle$, together with the prediction of Eq. \eqref{eq:rhoGGE}, $\textrm{Tr} \left[ \hat{m} \hat{\rho}_{\textrm{GGE}} \right]$, with $\beta=0.78219$, $\lambda_{c}=-3.39076$, $\lambda_{k}=-5.87297$ and $\lambda_{\pi}=0$. We thus confirm that symmetry breaking survives for long times (see below for a further numerical experiment involving much larger time scales), and that Eq. \eqref{eq:rhoGGE} provides a remarkable description, considering that our system only has $N=19$ particles.

The existence of this time-averaged symmetry-breaking state may be compatible with two scenarios: a coherent superposition of the two magnetization wells, corresponding to ES3, and a (classical) statistical mixture of these, corresponding to ES2. To distinguish between them, we study the instantaneous evolution of $\hat{W} = i \, \left| N \right\rangle \, \left\langle -N \right| - i \left| -N \right\rangle \, \left\langle N \right|$, where $\left| M\right>$ is the eigenstate of $\hat{M}$ with eigenvalue equal to $M$. If a symmetry-breaking state consists in a statistical mixture of states trapped in one of the two sides of the double well, then $\langle \hat{W}\rangle=0$. 
Hence, our numerical results show that our time-evolved state is in fact a \textit{coherent superposition of both magnetization wells}, as $\langle\hat{W}\rangle\neq 0$. Furthermore, the GGE prediction, displayed as a solid line, is in very good agreement with the numerical results.

\section{Symmetry-breaking perturbation}

Up to now, we have considered that the Hamiltonian is exactly given by Eq. \eqref{eq:TFIM}. However, real systems are usually affected by small perturbations which may entail significant consequences; indeed, such perturbations are proposed as the mechanism responsible of SSB. We finish our work by analyzing the effect of a symmetry-breaking perturbation in the Hamiltonian,

\begin{equation}\label{eq:HSB}
\hat{H}_{\epsilon}=    \hat{H}_{\textrm{TFIM}}+\frac{\epsilon}{2}\sum_{i}\hat{\sigma}_{i}^{x},
\end{equation}
where, typically, $|\epsilon| \ll 1$. 

The first consequence is that $[\hat{H}_{\epsilon},\hat{\Pi}]\neq 0$, so $\hat{\Pi}$ is no longer a conserved quantity if $\epsilon\neq 0$. Notwithstanding, as shown in Theorem 2.1, the double-well structure of the magnetization may survive within a certain subspace $\mathcal{H}_{\mathcal{C}}$; and, if $|\epsilon| \ll 1$, it is reasonable to expect that $\mathcal{H}_{\mathcal{C}} \sim \mathcal{H}_{SB}$. This entails that $\hat{\mathcal{C}}$ remains as the only conserved charge within $\mathcal{H}_{\mathcal{C}}$. Therefore, equilibrium states ES1 and ES2, i.e. with no quantum coherence, are the only possible ones under these circumstances.

\begin{figure*}[t!]
\centering
    \begin{tabular}{c c}
\hspace{-0.3cm}\includegraphics[width=0.52\textwidth]{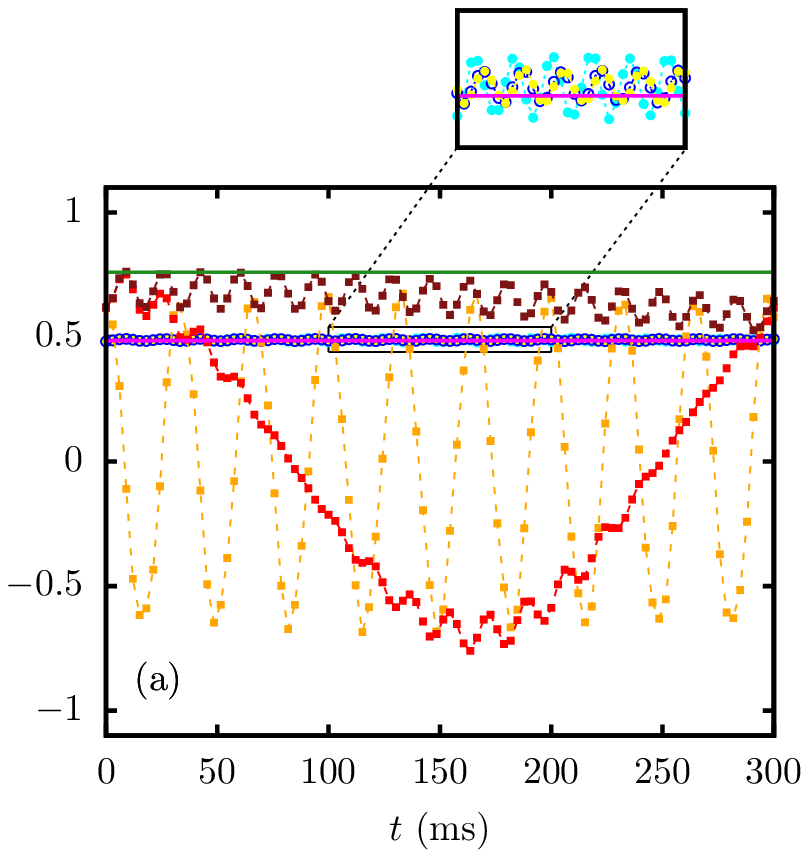} &
\hspace{-0.7cm}\includegraphics[width=0.52\textwidth]{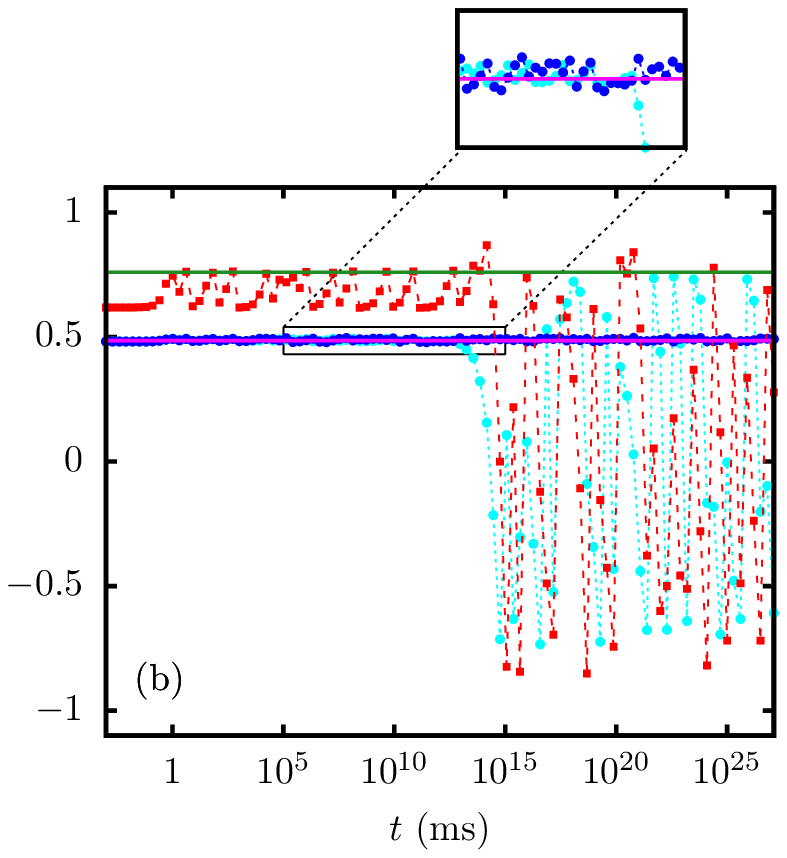}\\ 
    \end{tabular}
    \caption{\textbf{Dynamics in the symmetry-breaking TFIM.} (a) Instantaneous expectation value of the scaled magnetization $\hat{m}$ (circles) and $\hat{W}$ (squares) after a quench $h=0.5, \epsilon=0\to h=0.1,\epsilon\neq0$, $J=2$, with different symmetry-breaking strengths: for $\hat{m}$, $\epsilon=10^{-2}$ (cyan), $\epsilon=-10^{-3}$ (yellow) and $\epsilon=10^{-4}$ (blue, empty), while for $\hat{W}$, $\epsilon=10^{-2}$ (orange), $-10^{-3}$ (red) and $10^{-4}$ (brown). The initial state Eq. \eqref{eq:initialstate} has $p=1/2$ and $\phi=\pi/3$. The GGE predictions for $\hat{m}$ (magenta) and $\hat{W}$ (green) are also shown for the case $\epsilon=10^{-4}$. On the scale of the figure, $\hat{m}$ does not change significantly for the $\epsilon$ considered. (b) Instantaneous evolution of $\hat{m}$ (circles) for $\epsilon=0$ (cyan), $\epsilon=10^{-4}$ (blue) and $\hat{W}$ (squares) for $\epsilon=0$ (red). The magenta and green solid horizontal lines represent the GGE prediction for $\hat{m}$ ($\epsilon=10^{-4}$) and $\hat{W}$ ($\epsilon=0$), respectively. }
    \label{fig:symmetrybreaking}
\end{figure*}

To test this possibility, we come back to step S4 and consider the same state after this stage. Then, we perform a final quench from $h=0.5$ and $\epsilon=0$ to $h=0.1$ and different values of $\epsilon$. Results are displayed in Fig. \ref{fig:symmetrybreaking}(a). The first remarkable consequence is that $\left\langle \hat{m}(t) \right\rangle$ is the same for different symmetry-breaking strengths, $\epsilon$, {\em and signs}. 
Furthermore, the results for the three cases are well described by Eq. \eqref{eq:rhoGGE} with $\lambda_{\pi}=\lambda_k=0$ and $\lambda_{c}=-0.62158, -0.542119,-0.55005$ ($\epsilon=10^{-2},-10^{-3},10^{-4}$). 

These results confirm the scenario that motivated this set of numerical experiments. As a consequence of the small symmetry-breaking perturbation, $\hat{\mathcal{C}}$ remains as the only conserved charge within $\mathcal{H}_{\mathcal{C}}$. Therefore, its initial expectation value is conserved by the unitary time evolution, {\em irrespective of the sign of the perturbation, $\epsilon$}, and, after a sufficiently long-time evolution, we obtain an equilibrium state of type ES2, determined only by the energy and the initial value of $\langle \hat{\mathcal{C}} \rangle$. Notwithstanding, we note that the original equilibrium state, given by Eq. \eqref{eq:rhoGGE} with $\lambda_{c}=-3.39076$, $\lambda_{k}=-5.87297$ and $\lambda_{\pi}=0$ remains as a quite long-lived pre-thermal state \cite{Berges2004,Marcuzzi2013}, if $|\epsilon|$ is small enough. 

To delve into this last observation, we study the behavior at much larger time scales in Fig. \ref{fig:symmetrybreaking}(b). On the one hand, if $\epsilon=0$, we observe that symmetry-breaking is destroyed at very large time scales ---both $\langle \hat{m}(t) \rangle$ and $\langle \hat{W}(t) \rangle$ start to oscillate around zero at $\tau \sim 10^{15}$ ms. On the other hand, an infinitesimal perturbation, $\epsilon=10^{-4}$, stabilizes the symmetry-breaking state ---$\left\langle \hat{m} (t) \right\rangle$ follows the prediction of Eq. \eqref{eq:rhoGGE} with $\lambda_{\pi} = \lambda_k=0$ and $\lambda_{c}=-0.55005$ for the computed times. As explained, this is a consequence of Theorem 2.1. Note that, as we use a positive perturbation, $\epsilon>0$, the ground state of the system has $\left\langle \hat{m} \right\rangle < 0$. However, {\em as the main consequence of the perturbation is to stabilize the conservation of $\hat{\mathcal{C}}$, we obtain a symmetry-breaking equilibrium state with $\left\langle \hat{m} \right\rangle > 0$}, determined by the initial condition, even though we are working with a quite small system with $N=19$.

\section{Discussion}

We have proved that a discrete symmetry-breaking phase exists if and only if the quantum dynamics conserves three non-commuting charges. As a consequence of that, equilibrium states consisting in coherent superpositions of the different branches of the order parameter may be found. Furthermore, our theory also shows that the main consequence of introducing a small symmetry-breaking perturbation in the Hamiltonian is that only one of the three former charges remain constant, and that this fact stabilizes symmetry-breaking equilibrium states even in small systems.

Besides that, our main result also explains a number of experimental and theoretical results already obtained. The protocols used in Refs. \cite{Islam2013,Islam2011,Simon2011} give rise to equilibrium states with well defined parities, which belong to the family ES3 defined by Eq. \eqref{eq:rhoGGE}. The same kind of equilibrium states explains the bimodal structure for the order parameter numerically found in the long-range Ising model \cite{Ranabhat2022}. And the only conserved charge present when introducing a symmetry-breaking perturbation in the Hamiltonian provides an explanation for the failure of the eigenstate thermalization hypothesis \cite{Fratus2015,Mondaini2016}, and for the absence of thermalization after a quench \cite{Reimann2021} numerically found in such models. Finally, Theorems 1, 2 and 3 account for the observation of symmetry-breaking steady-states linked to excited-state \cite{Cejnar2021,Caprio2008,Puebla2013,Puebla2013b,Puebla2015} and dynamical quantum phase transitions \cite{Sciolla2011,Zhang2017,Muniz2020,Halimeh2017,Marino2022}, and provide a solid support for the characterization of different excited-state phases \cite{Corps2021} and the mechanism for dynamical quantum phase transitions in terms of conserved charges \cite{Corps2022,Corps2023}.

\section{acknowledgments}
 The authors thank R. Brito, J. M. R. Parrondo and P. P\'{e}rez-Fern\'{a}ndez for useful discussions. This work has been supported by the Spanish grant PGC-2018-094180-B-I00 and PID2019-106820RB-C21, funded by Ministerio de Ciencia e Innovaci\'{o}n/Agencia Estatal de Investigaci\'{o}n MCIN/AEI/10.13039/501100011033 and FEDER "A Way of Making Europe". A. L. C. acknowledges financial support from `la Caixa' Foundation (ID 100010434) through the fellowship LCF/BQ/DR21/11880024.

\newpage

~

\newpage

\onecolumngrid
\section{Supplemental Material}
Here we provide the proofs of Theorems 1-4 and technical details concerning our numerical simulations of the TFIM and the fitting procedure to obtain the generalized temperatures of the GGE. 

\subsection{Proofs of the theorems}

\textbf{Proof 1.-}  \textit{Forward implication.-} Consider an initial state given by a superposition of eigenstates of $\hat{H}$, $
\ket{\Psi(0)}=\sum_{n,\alpha} \psi_{n,\alpha}\ket{E_{n,\alpha}},\,\,\,n\in\mathbb{N},\,\,\alpha\in\{+,-\}.$ Its time evolution is $
\ket{\Psi(t)}=e^{-i\hat{H}t}\ket{\Psi(0)}=\sum_{n,\alpha}\psi_{n,\alpha}e^{-iE_{n,\alpha}t}\ket{E_{n,\alpha}},
$ where $\psi_{n,\alpha}=\bra{E_{n,\alpha}}\ket{\Psi(0)}\in\mathbb{C}$. For an operator $\hat{M}$ satisfying the property in \eqref{eq:propertyM}, the instantaneous expectation value is
\beq
\langle \hat{M}(t)\rangle=\sum_{n,m,\ell}\sum_{\alpha,\beta}c_{\ell} \psi_{m,\beta}^{*}\psi_{n,\alpha}e^{-i(E_{n,\alpha}-E_{m,\beta})t}\bra{E_{m,\beta}}\ket{E_{\ell,-\alpha}}.
\eeq
By virtue of the orthonormality of the eigenstate basis $\{\ket{E_{n,\alpha}}\}$, $\bra{E_{n,\alpha}}\ket{E_{m,\beta}}=\delta_{n,m}\delta_{\alpha,\beta}$, the only terms different from zero are those for which $m=\ell$ and $\beta=-\alpha$. Therefore, the long-time average of $\langle \hat{M}(t)\rangle$ is
\beq
\label{eq:promedio}
\begin{split}
\overline{\langle \hat{M}(t)\rangle}&=\sum_{n,m}\sum_{\alpha}c_{m}\psi_{m,-\alpha}^{*} \psi_{n,\alpha} \, \times \\ &\times \lim_{\tau \rightarrow \infty} \frac{1}{\tau} \int_0^{\tau} \, e^{-i(E_{n,\alpha}-E_{m,-\alpha})t} \, dt.
\end{split}
\eeq
The limit is zero except in those terms for which $E_{n,\alpha} = E_{m,-\alpha}$. So, the existence of an initial state for which $\overline{\langle \hat{M} \rangle} \neq 0$ implies the existence of such degenerate pairs. Without loss of generality, we consider that these pairs occur with $n=m$, so that 
\beq\label{eq:propertyE}
E_{n,\alpha}=E_{n,-\alpha},\,\,\,n\in D,\,\,\alpha\in\{+,-\},
\eeq
where $D=\{1, 2, \ldots, d\}$. Hence, the long-time average of $\langle \hat{M}(t) \rangle$ is
\beq
\label{eq:promedio2}
\overline{\langle \hat{M}(t) \rangle} = \sum_n \sum_{\alpha} c_n \psi^*_{n,-\alpha} \psi_{n,\alpha},\,\,\,n\in D,\,\,\alpha\in\{+,-\}.
\eeq
Since this summation can always be made different from zero by properly choosing the coefficients $\psi_{n,\alpha}$ of the initial state, we conclude that $\mathcal{H}_{SB}$ is a subspace of dimension $d$ spanned by all the eigenstate pairs, $\left\{ \ket{E_{n,+}}, \ket{E_{n,-}} \right\}$ with $n=1,2, \ldots, d$.

In this subspace, the original $\mathbb{Z}_2$ symmetry takes the form of a $2 \times 2$ block-diagonal matrix, each block spanned by a degenerate pair $\{ \ket{E_{n,+}}, \ket{E_{n,-}}\}$, given by
\beq
\hat{O}_{3}=\hat{\Pi}=\mqty(1 & 0 \\ 0 & -1).
\eeq

Consider now two operators, $\hat{O}_1$ and $\hat{O}_2$, having also the form of $2 \times 2$ block-diagonal matrices, each block given by
\beq\label{eq:matrixCK}
\hat{O}_{1}=\mqty(0 & 1 \\ 1 & 0), \,\,\,\hat{O}_{2}=\mqty(0 & -i \\ i & 0),
\eeq
in the same basis. From this structure it is obvious that the set $\left\{ \hat{O}_1, \hat{O}_2, \hat{O}_3 \right\}$ has the properties {\em (i)}, {\em (ii)} and {\em (iii)} of Theorem 1.

{\em Backward implication.-} Let us start by considering $\hat{O}_3 = \hat{\Pi}$, which is diagonal in the parity eigenbasis,
\beq
\hat{\Pi} \ket{E_{m,\alpha}}= \alpha \ket{E_{m,\alpha}}, \textrm{ with } \alpha= \pm 1.
\eeq
And let us consider also two other operators, $\hat{O}_1$ and $\hat{O}_2$, such that $\hat{O}_1^2 = \hat{O}_2^2=\mathbb{I}$, and the set $\left\{ \hat{O}_1, \hat{O}_2, \hat{O}_3 \right\}$ satisfies the SU(2) commutation rules. This last property requires that
\beq
\bra{E_{m,\alpha}} \hat{O}_1 \ket{E_{n,\alpha}} =\bra{E_{m,\alpha}} \hat{O}_2 \ket{E_{n,\alpha}} = 0, \; \forall n,m,\alpha.
\eeq
So, the only way that $\hat{O}_1$ and $\hat{O}_2$ can fulfill also the property {\em (iii)} of Theorem 1, that is, being constants of motion within $\mathcal{H}_{SB}$, is that $E_{n,\alpha}=E_{m,-\alpha}$ in all cases where $\bra{E_{m,\alpha}} \hat{O}_1 \ket{E_{n,-\alpha}} \neq 0$ or $\bra{E_{m,\alpha}} \hat{O}_2 \ket{E_{n,-\alpha}} \neq 0$. 

Now, let us consider an initial state, $\ket{\Psi(0)} = \sum_{m,\alpha} \psi_{m,\alpha} \ket{E_{m,\alpha}}$, with real coefficients satisfying $\psi_{m,\alpha} > 0$ if $c_m>0$, where the eigenstate $\ket{E_{m,\alpha}}$ belongs to a degenerate pair, and being zero in any other case. Then, from Eq. \eqref{eq:promedio2}, it is obvious that $\overline{\langle \hat{M} \rangle}>0$. 

Note that this argument also implies that $\hat{O}_1$ and $\hat{O}_2$ cannot commute with the projection of $\hat{H}$ onto $\mathcal{H}_D$, because this would imply the existence of initial states giving rise to $\overline{\langle \hat{M} \rangle}>0$ within this subspace.
$\blacksquare$

\underline{\textit{Comment}.-} Note that properties {\em (i)} and {\em (ii)} of Theorem 1 only apply to the projections of $\hat{O}_1$ and $\hat{O}_2$ onto $\mathcal{H}_{SB}$. The properties of these operators outside this subspace are not relevant.

\textbf{Proof 2.1.-} \textit{Forward implication.-} Consider the orthonormal basis formed by the eigenvectors of the observable $\hat{A}$, $\{\ket{A_{n}}\}_{n}$, where $\hat{A}\ket{A_{n}}=A_{n}\ket{A_{n}}$, and a state $\ket{\Psi(t)}=\sum_{n}  \psi_{n}(t)\ket{A_{n}}$. Consider also that the probability of measuring a positive eigenvalue of $\hat{{A}}$ is zero, i.e., $
\mathcal{P}(A>0)=\left|\left\langle A_{n}\right|\sum_{k} \psi_{k}(t)\left|A_{k}\right>\right|^{2}=0,\,\,\forall t,\,\,\,\forall\ket{A_{n}}\slash A_{n}>0.
$, which is satisfied if and only if $\psi_{n}(t)=0$, $\forall t$, if $A_{n}>0$.  Now, let us focus on $\langle \hat{\mathcal{C}} (t) \rangle$ for an initial state fulfilling the previous requirement. As $\hat{\mathcal{C}} = \textrm{sign} (\hat{A})$, we have that $\hat{\mathcal{C}} \ket{A_n} = \textrm{sign} \left( A_n \right) \ket{A_n}$. Therefore, as the only terms that contribute to the time evolution $\langle \hatmath{C}(t)\rangle$ are those coming from eigenvectors of negative $A_{n}$, we obtain
\beq
\langle \hat{\mathcal{C}}(t)\rangle=-\sum_{n,m} \psi_{m}^{*}(t)\psi_{n}(t)\delta_{m,n}=-\sum_{n}| \psi_{n}(t)|^{2}=-1,\,\,\,\forall t,
\eeq
which is constant. Alternatively, if $\mathcal{P}(A<0)=0$, then $\psi_{n}(t)=0$, $\forall t$, if $A_{n}<0$, and therefore the same calculation yields $\langle \hat{\mathcal{C}}(t)\rangle=+1$, $\forall t$.  

Next, let us focus on $\langle \hat{\mathcal{C}} (t) \rangle$ for an initial state given by a linear superposition of eigenstates of $\hat{A}$ whose eigenvalues have opposite sign,$
\ket{\Psi(0)}=\sum_{n,\,A_{n}<0}\psi_{n}(0)\ket{A_{n}}+\sum_{m,\,A_{n}>0}\psi'_{m}(0)\ket{A_{m}}$, 
The time evolution is now $
\ket{\Psi(t)}=\sum_{n,\,A_{n}<0}\psi_{n}(t)\ket{A_{n}}+\sum_{m,\,A_{n}>0}\psi'_{m}(t)\ket{A_{m}}.
$
The initial probability of measuring the eigenvalue $A_{n}<0$ ($A_{m}>0$) is $|\psi_{n}(0)|^{2}$ ($|\psi'_{m}(0)|^{2}$), so the total probability of obtaining any value $A<0$ ($A>0$) is $\sum_{n,\,A_{n}>0}|\psi_{n}(0)|^{2}$ ($\sum_{m,\,A_{m}<0}|\psi'_{m}(0)|^{2}$). Because the probability of measuring a value $A<0$ is conserved for all times by the Hamiltonian evolution, $
\sum_{n,\,A_{n}<0}|\psi_{n}(0)|^{2}=\sum_{n,\,A_{n}<0}|\psi_{n}(t)|^{2}
$ and similarly for $A>0$. Therefore, 
\beq
\langle \hat{\mathcal{C}}(t)\rangle=\sum_{n,\,A_{n}<0}|\psi_{n}(0)|^{2}-\sum_{m,\,A_{m}>0}|\psi'_{m}(0)|^{2},
\eeq
which is again constant, but not necessarily equal to $\pm 1$. Further, since the initial state is normalized, we find
$\langle \hat{\mathcal{C}}(t)\rangle=1-2\sum_{m,\,A_{m}>0}|\psi'_{m}(0)|^{2}\in[-1,1]$. 

\textit{Backward implication.-} If $\hat{\mathcal{C}}$ is a constant of motion, then the probabilities of observing its two eigenvalues, $\pm 1$, in a measurement are also constant. This implies that
\beq
\sum_{n,\,A_{n}<0}|\psi_{n}(t)|^{2} = \sum_{n,\,A_{n}<0}|\psi_{n}(0)|^{2}, \, \forall t,
\eeq
\beq
\sum_{n,\,A_{n}>0}|\psi_{n}(t)|^{2} = \sum_{n,\,A_{n}>0}|\psi_{n}(0)|^{2}, \, \forall t.
\eeq
Therefore, if $\mathcal{P}(A<0)=0$ (or $\mathcal{P}(A>0)=0$) in the initial state, then $\mathcal{P}(A<0)=0$ (or $\mathcal{P}(A>0)=0$) during the whole time evolution. $\blacksquare$ 

\textbf{Proof 2.2.}  By Theorem 2.1, the existence of $\hat{\Pi}$ is not necessary for the constancy of $\hat{\mathcal{C}}$. Yet, when $\hat{H}$ is invariant under $\hat{\Pi}$, the states that satisfy the conservation property of Theorem 2.1 must belong to $\mathcal{H}_{SB}$ of Theorem 1. Indeed, any state $\ket{\psi}\in\mathcal{H}_{D}=\mathcal{H}\setminus\mathcal{H}_{SB}$ is such that $\overline{\langle \hat{M}\rangle}=0$. These states can never verify that $\mathcal{P}(M>0)=0$ or $\mathcal{P}(M<0)=0$ for all time, since in that case $\overline{\langle \hat{M}\rangle}\neq 0$, and so $\ket{\psi}\in\mathcal{H}_{SB}$, which is a contradiction. Thus, $\hat{\mathcal{C}}$ is constant in $\mathcal{H}_{SB}$ only, so it commutes with the restriction of $\hat{H}$ to $\mathcal{H}_{SB}$. Finally, $\hat{\mathcal{K}}$ is conserved in $\mathcal{H}_{SB}$, because it is defined as the commutator of operators commuting in $\mathcal{H}_{\textrm{SB}}$.  $\blacksquare$\\

\underline{\textit{Comment}}: If $\hat{M}$ has at least one zero eigenvalue, then the operator $\textrm{sign}(\hat{M})$ is not well-defined. Yet, Theorem 2.2 holds true for $\hatmath{C}^{*}=\textrm{sign}^{*}(\hat{M})$ defined by

\beq
\hatmath{C}^{*}\ket{M_{n}}=
 \begin{cases}
      \ket{M_{n}}, & \text{if}\ M_{n}>0 \\
      -\ket{M_{n}}, & \text{if}\ M_{n}<0 \\ 0, & \text{if}\, M_{n}=0
    \end{cases}
\eeq
and for $\hatmath{K}^{*}=\frac{i}{2}[\hatmath{C}^{*},\hat{\Pi}]$.

\textbf{Proof 3.} Since $\hat{\Pi}$ is an Hermitian operator, any state can be decomposed in its eigenbasis as $\ket{\delta}=a\ket{\delta+}+b\ket{\delta-}$, where $\hat{\Pi}\ket{\delta \pm}=\pm\ket{\delta\pm}$.

To prove that the $\hat{\Pi}$, $\hatmath{C}$ and $\hatmath{K}$ operators satisfy the SU(2) commutation relations, we first note that since $\hat{\mathcal{C}}$ is a sign operator, it is immediately a $\mathbb{Z}_{2}$ operator, and thus $\hat{\mathcal{C}}^{2}=\mathbb{I}$. Parity is also trivially a $\mathbb{Z}_{2}$ operator. To show that $\hat{\mathcal{K}}^{2}=\mathbb{I}$, we will need to show that $\hat{\mathcal{C}}\hat{\Pi}\hat{\mathcal{C}}=-\hat{\Pi}$ and $\hat{\Pi}\hat{\mathcal{C}}\hat{\Pi}=-\hat{\mathcal{C}}$. For the first equality, we invoke the following integral representation of the sign operator \cite{Roberts1980,Denman1976}:

\beq
\hatmath{C}=\textrm{sign}\,(\hat{M})=\frac{2}{\pi}\hat{M}\int_{0}^{\infty}\textrm{d}t\,(t^{2}\mathbb{I}+\hat{M}^{2})^{-1}.
\eeq
It is clear that $\hatmath{C}$ is a parity-changing observable, since it is an odd function of $\hat{M}$, which inverts parity (note that the integrand is an even function of $\hat{M}$ and thus it conserves parity). Thus, we have 

\beq
\hat{\mathcal{C}}\hat{\Pi}\hat{\mathcal{C}}\ket{\delta}=\hat{\mathcal{C}}\hat{\Pi}\left(a\ket{\beta-}+b\ket{\beta+}\right)=\hat{\mathcal{C}}\left(-a\ket{\beta-}+b\ket{\beta+}\right).
\eeq
Because $\hat{\mathcal{C}}^{2}=\mathbb{I}$, it holds that $\hat{\mathcal{C}}^{2}\ket{\delta}=\ket{\delta}$, which means that $\hat{\mathcal{C}}\ket{\beta\pm}=\ket{\delta\mp}$. Therefore, 

\beq
\hat{\mathcal{C}}\hat{\Pi}\hat{\mathcal{C}}\ket{\delta}=-a\ket{\delta+}+b\ket{\delta-}=-\hat{\Pi}\ket{\delta},\,\,\,\forall \ket{\delta}\in\mathcal{H}.
\eeq
The second equality can be proved similarly. It then follows that $\hatmath{K}^{2}=\mathbb{I}$. 
By definition, $
[\hat{\mathcal{C}},\hat{\Pi}]=-2i\hat{\mathcal{K}}.$ Finally,$
[\hat{\mathcal{C}},\hat{\mathcal{K}}]=\frac{i}{2}(\hat{\mathcal{C}}^{2}\hat{\Pi}+\hat{\Pi}\hat{\mathcal{C}}^{2}-2\hat{\mathcal{C}}\hat{\Pi}\hat{\mathcal{C}})=2i\hat{\Pi},
$ and $
[\hat{\Pi},\hat{\mathcal{K}}]=\frac{i}{2}\left(-\hat{\Pi}^{2}\hat{\mathcal{C}}-\hat{\mathcal{C}}\hat{\Pi}^{2}+2\hat{\Pi}\hat{\mathcal{C}}\hat{\Pi}\right)=-2i\hatmath{C}$ $\blacksquare$

\underline{\textit{Comment.-}} Proof 3 is given for the case in which $\hat{M}$ has no zero eigenvalues. Consider now that this is not the case, so that there exists a subspace $\mathcal{M}_0 \neq \lbrace \emptyset \rbrace$ spanned by the eigenstates of $\hat{M}$ with zero eigenvalue. Then, the arguments used in proof 3 are only valid for $\mathcal{H} \backslash \mathcal{M}_0$. This means that, in this case, the operators $\hat{\mathcal{C}^*}$, $\hat{\mathcal{K}^*}$ and $\hat{\Pi}$ only satisfy properties (i) and (ii) of Theorem 1 within $\mathcal{H} \backslash \mathcal{M}_0$. Nevertheless, this problem has almost no practical relevance. As we have pointed out in the comment of proof 1, properties (i) and (ii) of Theorem 1 are relevant only within $\mathcal{H}_{SB}$. Hence, the key point is whether $\mathcal{H}_{SB} \cap \mathcal{M}_0$ is important or not. As in typical symmetry-breaking phases there exists a sort of {\em potential barrier} between the two wells of the order parameter, it is reasonable to expect that $\mathcal{H}_{SB} \cap \mathcal{M}_0 = \lbrace \emptyset \rbrace$ under physically realistic conditions. Therefore, the possible existence of the eigenvalue $M=0$ for the order parameter has no physical relevance. 

\textbf{Proof 4.}  Theorem 1 teaches us that within each subspace of $\mathcal{H}_{SB}$ spanned by the eigenvectors $\{\ket{E_{n,+}},\ket{E_{n,-}}\}$ with $E_{n,+}=E_{n,-}$ ($E_{n}<E_{c}$), $\hat{\mathcal{C}}$ and $\hatmath{K}$ take the form of \eqref{eq:matrixCK}. Therefore, in $\mathcal{H}_{SB}$ the combination $\hat{R}=\beta\hat{H}+\beta_{c}\hatmath{C}+\beta_{k}\hatmath{K}+\beta_{\pi}\hat{\Pi}$ is a block diagonal matrix of $2\times 2$ matrices in the eigenbasis common to parity and the Hamiltonian. For an $n$-dimensional $\mathcal{H}_{SB}$, this is
\beq
\hat{R}_{SB}=\textrm{diag} \{R_{i}\}_{i=1}^{n},\,\,\,R_{i}=\mqty(\beta E_{n,+}+\beta_{\pi} & \beta_{c}-i\beta_{k} \\ \beta_{c}+i\beta_{k} & \beta E_{n,-}-\beta_{\pi}). 
\eeq
Within $\mathcal{H}_{D}$, where $E_{n,+}\neq E_{n,-}$, the precise form of $\hat{R}_{D}$ is unknown. In the total Hilbert space $\mathcal{H}$, 

\beq
\hat{R}=\mqty(\hat{R}_{SB} & 0\\ 0 & \hat{R}_{D}),
\eeq
where $\hat{R}_{SB}$ is a matrix of order $\textrm{dim}(\mathcal{H}_{SB})$ and $\hat{R}_{D}$ is a matrix of order $\textrm{dim}(\mathcal{H}_{D})$. The exponential matrix of $\hat{R}$ must necessarily have the same structure. Therefore, we can build the following matrix
\beq
\hat{D}=\frac{e^{\hat{R}}}{Z} \equiv \frac{1}{Z} \mqty(\hat{T}_{SB} & 0 \\ 0 & \hat{T}_{D}),
\eeq
where $Z=\textrm{Tr}[e^{\hat{R}}]$ is a normalization constant. As $\hat{R}$ is hermitian, $\hat{D}$ is a positive-definite matrix with $\textrm{Tr}[\hat{D}]=1$, due to the normalization constant $Z$. Therefore, it is a density matrix, and all its eigenvalues are $d_{n}> 0$. 

Let us assume now that the canonical and microcanonical ensemble are equivalent, so that $\sigma_E/E \rightarrow 0$ in the TL,
as stated in hypothesis (H4). This means that if $\beta>\beta_{c}$ ($T<T_{c}$), the probability of populating a state beyond $E_{c}$ becomes zero in the TL, and therefore, $\Tr\,\hat{T}_{D} \rightarrow 0$ in the TL. Furthermore since $\hat{D}$ is definite positive, all eigenvalues of $\hat{T}_{D}$ must remain positive for any finite system size, becoming zero in the TL. And, because $\hat{D}=\hat{D}^{\dagger}$, this further implies that $\hat{T}_{D} \rightarrow 0$, and that $\Tr\,\hat{T}_{SB} \rightarrow 1$ in the TL. Finally, as $\hat{T}_{SB}$ is a block diagonal matrix composed of $2\times 2$ matrices, and for $E<E_{c}$ the eigenvalues of opposite parity become degenerate in the TL, this means that $[\hat{H},\hat{D}] \rightarrow 0$ in the TL. $\blacksquare$

\subsection{Numerical simulation of the TFIM}
Here we give some technical details of our numerical simulation of the TFIM \eqref{eq:TFIM}. We choose power-law long-range interactions controlled by a parameter $\alpha\in[0,\infty)$, with $\alpha=0$ denoting the fully-connected limit and $\alpha\to\infty$ corresponding to the nearest-neighbor TFIM. For our experiments, we choose $\alpha=1.1$. The Hamiltonian \eqref{eq:TFIM} is invariant under the inversion symmetry $\hat{\mathcal{I}}$. In the site basis, where each state is represented by the tensor product of the orientation of the $i$th spin ($i=1,2,\ldots,N$) along the $z$-axis, $\ket{\phi}=\bigotimes_{i=1}^{N}\ket{\phi_{i}}_{z}$, $\phi_{i}\in\{\uparrow,\downarrow\}$, this operator implements a reflection along the center of the chain, i.e., $\hatmath{I}\ket{\phi_{1}\,\phi_{2}\,\ldots\,\phi_{N}}_{z}=\ket{\phi_{N}\,\ldots\,\phi_{2}\,\phi_{1}}_{z}$. We only keep states $\ket{\phi}$ in the positive inversion sector, such that $\bra{\phi}\hatmath{I}\ket{\phi}=1$. We also implement periodic boundary conditions (PBCs). This implies that the Hamiltonian is also invariant under the translation symmetry, $\hatmath{T}\ket{\phi_{1}\,\phi_{2}\,\cdots,\phi_{N}}_{z}=\ket{\phi_{N}\,\phi_{1}\,\cdots,\phi_{N-1}}_{z}$. So, in a similar way than before, we only keep states $\ket{\phi}$ such that $\bra{\phi}\hatmath{T}\ket{\phi}=1$, which
is equivalent to working with a zero momentum basis. Finally, \eqref{eq:TFIM} also commutes with the parity $\hat{\Pi}=\prod_{i=1}^{N}\hat{\sigma}_{i}^{z}$ \cite{Russomanno2021}. We work with both parity sectors simultaneously. 

With our choice of PBCs, the interaction potential of \eqref{eq:TFIM} takes the form $V_{ij}=\frac{J}{\mathcal{N}(\alpha)}D_{ij}^{-\alpha}$, where $D_{ij}=\frac{1}{\min\{|i-j|,N-|i-j|\}}$ and $\mathcal{N}(\alpha)=\frac{1}{N-1}\sum_{i\neq j}D_{ij}^{-\alpha}$ being the Kac factor \cite{Kac1963}, which ensures the Hamiltonian intensiveness when $\alpha<1$ but could be omitted with $\alpha\geq 1$. Here, $J$ is an arbitrary coupling constant. 

As for the symmetry-breaking TFIM \eqref{eq:HSB}, the states do not have a definite parity quantum number, but we still work with the positive inversion and translation states, as described above.

\subsection{Fitting the GGE}

The instantaneous evolution the observables shown in this work are compared against the GGE prediction \eqref{eq:rhoGGE}. For a state in the symmetry-breaking phase of the TFIM \eqref{eq:TFIM}, the temperatures $\lambda_{c}$, $\lambda_{k}$ and $\lambda_{\pi}$ are fixed by the initial condition and may be obtained by solving the following system of non-linear equations 

\begin{spreadlines}{1ex}
\begin{equation}
\begin{dcases}
2p-1=-\lambda_{\pi}\frac{\tanh\sqrt{\lambda_{\pi}^{2}+\lambda_{c}^{2}+\lambda_{k}^{2}}}{\sqrt{\lambda_{\pi}^{2}+\lambda_{c}^{2}+\lambda_{k}^{2}}},\\
2\sqrt{p(1-p)}\cos\phi=-\lambda_{c}\frac{\tanh\sqrt{\lambda_{\pi}^{2}+\lambda_{c}^{2}+\lambda_{k}^{2}}}{\sqrt{\lambda_{\pi}^{2}+\lambda_{c}^{2}+\lambda_{k}^{2}}}, \\
2\sqrt{p(1-p)}\sin\phi=-\lambda_{k}\frac{\tanh\sqrt{\lambda_{\pi}^{2}+\lambda_{c}^{2}+\lambda_{k}^{2}}}{\sqrt{\lambda_{\pi}^{2}+\lambda_{c}^{2}+\lambda_{k}^{2}}},
\end{dcases}
\end{equation}
\end{spreadlines}
which results from imposing that the GGE yield the corresponding value for each of the non-commuting charges, $\textrm{Tr}\,[\hat{\rho}_{\textrm{GGE}}\hatmath{O}]=\langle \hatmath{O}\rangle$ with $\hatmath{O}=\hatmath{C},\hatmath{K},\hat{\Pi}$. The remaining temperature $\beta$ is then obtained by a single fit reproducing the final (average) energy of the quenched state, $\textrm{Tr}\,[\hat{\rho}_{\textrm{GGE}}\hat{H}]=\langle E\rangle$. For the symmetry-breaking Hamiltonian \eqref{eq:HSB}, $\lambda_{k}=\lambda_{\pi}=0$; $\lambda_{c}$ and $\beta$ are obtained through two fits to the value of $\hatmath{C}$ and the final energy, respectively. Overall, we observe a good agreement of the exact results and the GGE. The small discrepancies are finite-size effects of the TFIM with only $N=19$ particles.

\end{document}